# Black-body Thermal Radiative and Thermodynamic Functions of 1-Dimensional Self-Assembly Nanotubes

Anatoliy I Fisenko, Vladimir F Lemberg

*ONCFEC Inc., 250 Lake Street, Suite 909, St. Catharines, Ontario L2R 5Z4, Canada*

*E-mail: afisenko@oncfec.com*

**Abstract** Assuming that *1* - dimensional self-assembly nanotubes can be represented as a *1*-dimensional cavity-type black-body radiator of length *L* and radius *r* in thermal equilibrium at a temperature *T*, analytical expressions for the thermal radiative and thermodynamic functions of the emitted black-body radiation are obtained in the finite spectral range of the electromagnetic spectrum. The total energy density, the Stefan-Boltzmann law, the number density of the photons, the Helmholtz free energy density, the entropy density, the heat capacity at constant volume, and the pressure are expressed in terms of the polylogarithm functions. In the frequency range 0.05 – 0.19 PHz, the thermal radiative and thermodynamic functions of the *1* - dimensional black-body radiation at different temperatures are calculated. To confirm our assumption, it is necessary to measure the total energy density in the direction *L* in the finite frequency range 0.05 - 0.19 PHz, using a single nanoparticle mass spectrometer (SNMS) or other optical devices.

## 1. Introduction

It is well-known that some materials at nanoscale ranges have the unique ability to self-assemble into organized structures creating *1, 2* and *3* dimension systems [1, 2, 3].

One-dimensional nanomaterials (*1*D) have one dimension that is outside the nanoscale (<100 nm) range. This means that there are two dimensions are at the nanoscale and one dimension at the microscale. Typical examples of one-dimensional nanomaterials are nanotubes, nanorods, and nanowires. Unlike bulk materials, reducing the size of the materials to the nanoscale range will affect the optical, electrical, and magnetic behavior of the materials.

The field of *1*-dimensional nanostructured materials, such as nanotubes, in recent years has attracted considerable attention of many researchers because of their simplicity in preparation and improved catalytic properties compared to bulk materials [3, 4, 5]. These materials include *1*-dimensional carbon nanotubes. A unique feature of carbon nanotubes is used for the manufacture of LEDs [6, 7] and photodetectors [8].

In [9], the thermal radiative and thermodynamic properties of *1*-dimensional blackbody radiation in a semi-infinite frequency range ($0 \leq v \leq \infty$) were studied. However, as was noted in [10–13] for practical applications, thermal radiative and thermodynamic functions must be obtained in a specific finite range of the spectrum of thermal radiation. Currently, there are no detailed studies in the existing literature related to the construction of thermal radiative and thermodynamic functions of *1*-dimensional blackbody radiation in a finite frequency range (wavelength, wave number) of the spectrum ($v_1 \leq v \leq v_2$).

This paper is devoted to a detailed study of the thermal radiative and thermodynamic functions of *1*- dimensional black-body radiation in a finite range of a spectrum at various temperatures. Analytical expressions for the total energy density, the Stefan-Boltzmann law, the number density of the photons, the Helmholtz free energy density, the entropy density, the heat capacity at constant volume, and the pressure are obtained. The thermal radiative and thermodynamic functions of *1*-dimensional black-body radiation are calculated in the frequency range 0.05 – 0.19 PHz at various temperatures. The possibility of applying the obtained results to measurements of *1*-dimensional thermal radiative and thermodynamic functions for carbon nanotubes is discussed.

## 2. General expressions of the thermal radiative and thermodynamic functions of *1*-dimensional black-body radiation in specific spectral ranges

Let us consider a *1*-dimensional cavity-type black-body radiator of length $L$ and radius $r$ ($L \gg r$) in thermal equilibrium at temperature $T$. Carbon nanotubes are as examples of an unique *1*-dimensional radiator which can be envisioned as rolled single sheets of graphite or graphene. The diameter typically varies in the range 0.4–40 nm, but the length can vary from 0.14 nm to 55.5 cm [14].

In [15, 16], it was shown that a carbon nanotube itself emits thermal radiation in *3*-D space very close to a black-body. However, a carbon nanotube has a hollow nanostructure with *1*-D free space inside. Therefore, it is interesting to study the thermal radiative and thermodynamic properties of black-body radiation emitted from the *1* - dimensional black-body radiator at various temperatures. In this case, to obtain expressions for the thermal radiative and thermodynamic functions, we should use the *1*-dimensional Planck formula for blackbody radiation.

Below, the thermal radiative and thermodynamic functions of black-body radiation emitted from of *1*-dimensional cavity-type black-body radiator are obtain. To confirm our results, measurements using a single nanoparticle mass spectrometer (SNMS) [17] or other optical devices are necessary.

### 2.1 Partial Thermal radiative functions of *1*-dimensional black-body radiation

The spectral energy density of *1*-dimensional black-body radiation at temperature $T$ is given by the generalized Planck radiation law [9, 18]

$$I_P^1(\nu, T) = \frac{4h}{c} \frac{\nu}{e^{\frac{h\nu}{k_B T}} - 1}, \qquad (1)$$

where $T$ denotes the black-body temperature, $h$ and $k_B$ are the Planck and Boltzmann constants, respectively. $c$ is the speed of light.

According to Eq. (1), the total energy density over a finite range of frequencies is defined as

$$I^1(\nu_1, \nu_2, T) = \frac{4h}{c} \int_{\nu_1}^{\nu_2} \frac{\nu}{e^{\frac{h\nu}{k_B T}} - 1} d\nu . \qquad (2)$$

By computing the integral in Eq. (2), we obtain

$$I^1(x_1, x_2, T) = \frac{4k_B^2}{ch} T^2 A(x_1, x_2) = a^1 T^2, \qquad (3)$$

where

$$A(x_1, x_2) = [P_1(x_1) - P_1(x_2)]. \qquad (4)$$

Here $x = \frac{h\nu}{k_B T}$. $P_n(x)$ is defined as

$$P_n(x) = \sum_{s=0}^{n} \frac{(x)^s}{s!} \text{Li}_{n+1-s}(e^{-x}), \qquad (5)$$

where

$$\text{Li}_{n+1-s}(e^{-x}) = \sum_{k=1}^{\infty} \frac{e^{-kx}}{k^{n+1-s}} \qquad (6)$$

is the polylogarithm functions [19].

The constant $a^1 = \frac{4k_B^2}{ch} A(x_1, x_2)$ in Eq. (3) is the *1*- dimensional radiation density constant in a finite range from $x_1$ to $x_2$. In the semi-infinite range, we have $P_1(0) = \text{Li}_2(1) = \xi(2) = \frac{\pi^2}{6}$ and $P_1(\infty) = 0$, then Eq. (3) transform in $I^1(0, \infty) = \frac{2k_B^2 \pi^2}{3ch} T^2$. Here $a^1 = \frac{2k_B^2 \pi^2}{3ch} = 6.314 \times 10^{-21} \frac{\text{J}}{\text{m K}^2}$ is the *1*-dimensional radiation density constant in the semi-infinite range of frequency.

The generalized *1*-dimensional Stefan-Boltzmann law takes the form,

$$I^{SB\,1}(x_1, x_2, T) = \frac{k_B^2}{h} T^2 A(x_1, x_2), \qquad (7)$$

where $\sigma^1 = \frac{k_B^2}{h} A(x_1, x_2)$ is the *1*- dimensional Stefan-Boltzmann constant in a finite range. In the semi-infinite range, the expression for $\sigma^1$ takes the form, $\sigma^1 = \frac{k_B^2 \pi^2}{6h} = 9.64 \times 10^{-13} \frac{\text{W}}{\text{K}^2}$. Eq. (7) can be used when considering *1*-dimensional heat transfer with radiation along a nanotube. Unlike *3*-dimensional materials, the generalized *1*-dimensional Stefan-Boltzmann law depends on temperature to the second degree.

According to Eq. (1), the number density of photons can be written as [9]:

$$n^1(v_1, v_2, T) = \frac{4}{c} \int_{v_1}^{v_2} \frac{dv}{e^{\frac{hv}{k_B T}} - 1} \quad . \tag{8}$$

By computing the integral in Eq. (8), we have

$$n^1(x_1, x_2, T) = \frac{4k_B}{hc} T [P_0(x_1) - P_0(x_2)], \tag{9}$$

where

$$B(x_1, x_2) = [P_0(x_1) - P_0(x_2)]. \tag{10}$$

## 2. 2 Thermodynamics of *1*-dimensional black-body radiation in a finite range of frequency

Let us now construct the thermodynamics of *1*-dimensional black-body radiation in a finite range of frequency.

(1) Helmholtz free energy density: $f^1 = \frac{F^1}{L}$.

Helmholtz free energy density can be presented in the form [9],

$$f^1(v_1, v_2, T) = \frac{4}{c} T \int_{v_1}^{v_2} \ln\left(1 - e^{-\frac{hv}{k_b T}}\right) dv. \tag{11}$$

By computing the integral in Eq. (11), we obtain

$$f^1(x_1, x_2, T) = -\frac{4k_B^2}{ch} T^2 \, C(x_1, x_2), \tag{12}$$

where

$$C(x_1, x_2) = \left[(P_1(x_1) - P_1(x_2)) - (x_1 \mathrm{Li}_1(e^{-x_1}) - x_2 \, \mathrm{Li}_1(e^{-x_2}))\right]. \tag{13}$$

In the semi-infinite range of *x*, we obtain: $f^1(0, \infty, T) = -\frac{2k_B^2 \pi^2}{3ch} T^2$.

(2) Entropy density: $s^1 = \frac{S^1}{L} = -\frac{\partial f^1}{\partial T}$.

The entropy density has the following structure:

$$s^1(x_1, x_2, T) = \frac{8k_B^2}{ch} T \, D(x_1, x_2), \tag{14}$$

where

$$D(x_1,x_2) = \left[ (P_1(x_1) - P_1(x_2)) - \frac{1}{2}(x_1 \text{Li}_1(e^{-x_1}) - x_2 \text{Li}_1(e^{-x_2})) \right]. \tag{15}$$

In the semi-infinite range of $x$, $P_1(0) = \text{Li}_2(1) = \xi(2) = \dfrac{\pi^2}{6}$ and $P_1(\infty) = 0$, then

$$s^1(0,\infty) = \frac{4k_B^2 \pi^2}{3ch} T \ .$$

(3) Heat capacity at constant volume per unit volume: $c^1{}_V = -\dfrac{\partial I^1}{\partial T}$.

For the heat capacity at constant volume per unit volume, we have

$$c_V^1(x_1, x_2, T) = \frac{8k_B^2}{ch} T\, E(x_1, x_2), \tag{16}$$

where

$$E(x_1,x_2) = \left[ (P_1(x_1) - P_1(x_2)) + \frac{1}{2}(x_1^2 \text{Li}_0(e^{-x_1}) - x_2^2 \text{Li}_0(e^{-x_2})) \right]. \tag{17}$$

In the semi-infinite range of $x$, $P_1(0) = \text{Li}_2(1) = \xi(2) = \dfrac{\pi^2}{6}$ and $P_1(\infty) = 0$, then

$$c_V^1(0,\infty) = \frac{4k_B^2 \pi^2}{3ch} T \ .$$

(4) Pressure: $p^1 = -f^1$.

$$p^1(x_1, x_2, T) = \frac{4k_B}{ch} T^2 C(x_1, x_2). \tag{18}$$

## 3. Results and Discussion

The calculated values of $A(x_1, x_2)$, $B(x_1, x_2)$, $C(x_1, x_2)$, $D(x_1, x_2)$, $E(x_1, x_2)$ as a function of $x_1$ and $x_2$ in the general expressions for thermal radiative and thermodynamic functions are presented in Table 1. The range of values $x_1$ and $x_2$ was chosen to exclude only the length of the manuscript.

In Table 2, Table 3, and Table 4, the calculated values of thermal radiative and thermodynamic functions of *1* and *3*-dimensional black-body radiation are presented in the frequency range 0.05 – 0.19 PHz at $T_1 = 298$ K, $T_2 = 1273$ K, and $T_3 = 2273$ K. This frequency range was chosen for calculation, since a single nanoparticle mass spectrometer uses this interval for measurement.

As can be seen from Tables 2-4, the values for *3*-dimensional thermal radiative and thermodynamic functions are significantly higher than for *1*- dimensional functions. For example, the *3*-dimensional value for the total energy density at a temperature $T = 2273$ K is $I^3(v_1,v_2,T) = 1.13 \times 10^{-2}$ J m$^{-3}$ and 12 orders of magnitude higher than for the *1*-dimentional value: $I^1(v_1,v_2,T) = 1.47 \times 10^{-14}$ J m$^{-1}$. Similar situations exist for other thermal radiative and thermodynamic functions. With increasing temperature the *1*-dimentional thermal radiative and thermodynamic functions increase.

According to Table 4, the *1*-dimensional calculated value for the total energy density at a temperature $T = 2273$ K is $I^1(v_1,v_2,T) = 1.47 \times 10^{-14}$ J m$^{-1}$. To confirm our result on carbon nanotubes, it is necessary to measure the total energy density in the direction $L$ in a finite frequency range from $v_1 = 0.05$ PHz to $v_2 = 0.19$ PHz at $T = 2273$ K using a single nanoparticle mass spectrometer (SNMS) or other optical devices.

It should be noted that the obtained analytical expressions for the thermal radiative and thermodynamic functions of the *1*-dimensional black-body radiation can be easily transformed to the wavenumber ($\tilde{v}$) or wavelength ($\lambda$) domains. In these cases, we should use the following relationships:

$$v = \frac{c}{\lambda} \quad , \quad v = c\tilde{v} \tag{19}$$

$$dv = -\frac{c}{\lambda^2} d\lambda \quad , \quad dv = cd\tilde{v} \tag{20}$$

The arbitrary variable $x$ in the expressions for thermal radiative and thermodynamic functions of the *1*-dimensional black-body radiation can be represented in all three domains as follows:

$$x = \frac{hv}{k_B T} = \frac{hc}{\lambda k_B T} = \frac{h\tilde{v}}{k_B T}. \tag{21}$$

The relationship between integrals in the finite spectral range for different domains is defined as:

$$\int_{v_1}^{v_2} I^1(v,T)\,\mathrm{d}v = -\int_{\lambda_1}^{\lambda_2} I^1(\lambda,T)\,\mathrm{d}\lambda = \int_{\tilde{v}_1}^{\tilde{v}_2} I^1(\tilde{v},T)\,\mathrm{d}\tilde{v} = I(x_1,x_2,T). \qquad (22)$$

As seen, the results of these integrals are identical because any variable substitution does not affect the value of the calculated integral. Thus, the solution is independent on the choice of the domain. Therefore, all analytical expressions for thermal radiative and thermodynamic functions of the *1*-dimensional black-body radiation, obtained above, have the same structure for the wavenumber and wavelength domains.

Note that using different domains produce the same results, because it represents the same physical quantity. This means that the calculated values presented in Tables 2-4 are valid for different domains.

## 4. Conclusions

The analytical expressions for the thermal radiative and thermodynamic functions of the *1*-dimensional black-body radiation emitted by cavity-type black-body radiator are obtained in a finite range of frequencies at different temperatures. The calculated values of $A(x_1,x_2)$, $B(x_1,x_2)$, $C(x_1,x_2)$, $D(x_1,x_2)$, $E(x_1,x_2)$ at different $x_1$ and $x_2$ in general expressions for the thermal radiative and thermodynamic functions are presented in Table 1. The total energy density, the Stefan-Boltzmann law, the number density of the photons, the Helmholtz free energy density, the entropy density, the heat capacity at constant volume, and the pressure of the *1*- dimensional blackbody radiation are calculated in a finite range of frequencies from $v_1 = 0.05$ PHz to $v_2 = 0.19$ PHz at $T_1 = 298$ K, $T_2 = 1273$ K, and $T_3 = 2273$ K and the results are presented in Tables 2-4. A comparative analysis of these functions with three-dimensional blackbody radiation is carried out.

The possibility of measuring the calculated values of the thermal radiative and thermodynamic functions of *1*- dimensional black-body radiation for carbon nanotubes using optical devices is indicated.

In conclusion, it is important to note that the results obtained in this article can be used when considering black-body radiation from any materials with hollow *1*- dimensional space as well as with hollow nanostructured metals, such as a novel class of plasmonic nanostructures.

There are several classes of *2*-dimensional nanomaterials for which it seems desirable to construct the thermal radiative and thermodynamic functions of *2*-dimensional blackbody radiation. These *2*-dimensional nanomaterials are: a) graphene; b) nanofilms; c) nanolayers; and d) nanocoatings.

These and other topics will be points of discussion in subsequent publications.

| $x_1$ | $x_2$ | $A(x_1,x_2)$ | $B(x_1,x_2)$ | $C(x_1,x_2)$ | $D(x_1,x_2)$ | $E(x_1,x_2)$ |
|---|---|---|---|---|---|---|
| 0.1 | 1.1 | 0.7365 | 0.2042 | 0.9465 | 0.8415 | 0.4822 |
| 0.1 | 2.1 | 1.1467 | 0.5217 | 1.1858 | 1.1662 | 0.8865 |
| 0.1 | 3.1 | 1.3589 | 0.7916 | 1.2666 | 1.3128 | 1.1798 |
| 0.1 | 4.1 | 1.4622 | 0.9743 | 1.2955 | 1.3789 | 1.3681 |
| 0.1 | 5.1 | 1.5101 | 1.0829 | 1.3061 | 1.4081 | 1.4779 |
| 0.1 | 6.1 | 1.5315 | 1.1419 | 1.3099 | 1.4207 | 1.5372 |
| 0.1 | 7.1 | 1.5407 | 1.1721 | 1.3114 | 1.426 | 1.5674 |
| 0.1 | 8.1 | 1.5446 | 1.1869 | 1.3119 | 1.4283 | 1.5822 |
| 0.1 | 9.1 | 1.5463 | 1.1939 | 1.3121 | 1.4292 | 1.5892 |
| 1.1 | 2.1 | 0.4102 | 0.3175 | 0.2392 | 0.3247 | 0.4043 |
| 1.1 | 3.1 | 0.6224 | 0.5874 | 0.3201 | 0.4713 | 0.6976 |
| 1.1 | 4.1 | 0.7257 | 0.7701 | 0.349 | 0.5374 | 0.886 |
| 1.1 | 5.1 | 0.7736 | 0.8787 | 0.3595 | 0.5666 | 0.9957 |
| 1.1 | 6.1 | 0.795 | 0.9377 | 0.3634 | 0.5792 | 1.055 |
| 1.1 | 7.1 | 0.8042 | 0.9679 | 0.3648 | 0.5845 | 1.0853 |
| 1.1 | 8.1 | 0.8081 | 0.9827 | 0.3653 | 0.5867 | 1.1 |
| 1.1 | 9.1 | 0.8098 | 0.9897 | 0.3655 | 0.5877 | 1.107 |
| 2.1 | 3.1 | 0.2123 | 0.2699 | 0.0809 | 0.1466 | 0.2933 |
| 2.1 | 4.1 | 0.3156 | 0.4526 | 0.1098 | 0.2127 | 0.4816 |
| 2.1 | 5.1 | 0.3635 | 0.5612 | 0.1203 | 0.2419 | 0.5914 |
| 2.1 | 6.1 | 0.3848 | 0.6202 | 0.1242 | 0.2545 | 0.6507 |
| 2.1 | 7.1 | 0.3941 | 0.6504 | 0.1256 | 0.2598 | 0.6809 |
| 2.1 | 8.1 | 0.398 | 0.6652 | 0.1261 | 0.2621 | 0.6957 |
| 2.1 | 9.1 | 0.3996 | 0.6722 | 0.1263 | 0.263 | 0.7027 |
| 3.1 | 4.1 | 0.1033 | 0.1828 | 0.0289 | 0.0661 | 0.1883 |
| 3.1 | 5.1 | 0.1512 | 0.2913 | 0.0395 | 0.0953 | 0.2981 |
| 3.1 | 6.1 | 0.1725 | 0.3503 | 0.0433 | 0.1079 | 0.3574 |
| 3.1 | 7.1 | 0.1818 | 0.3806 | 0.0447 | 0.1133 | 0.3876 |
| 3.1 | 8.1 | 0.1857 | 0.3953 | 0.0453 | 0.1155 | 0.4024 |
| 3.1 | 9.1 | 0.1873 | 0.4023 | 0.0455 | 0.1164 | 0.4094 |
| 4.1 | 5.1 | 0.0479 | 0.1085 | 0.0105 | 0.0292 | 0.1097 |
| 4.1 | 6.1 | 0.0692 | 0.1676 | 0.0144 | 0.0418 | 0.169 |
| 4.1 | 7.1 | 0.0785 | 0.1978 | 0.0158 | 0.0471 | 0.1993 |
| 4.1 | 8.1 | 0.0824 | 0.2126 | 0.0163 | 0.0494 | 0.2141 |
| 4.1 | 9.1 | 0.084 | 0.2195 | 0.0165 | 0.0503 | 0.221 |
| 5.1 | 6.1 | 0.0214 | 0.0591 | 0.0039 | 0.0126 | 0.0593 |
| 5.1 | 7.1 | 0.0306 | 0.0893 | 0.0053 | 0.0179 | 0.0896 |
| 5.1 | 8.1 | 0.0345 | 0.1041 | 0.0058 | 0.0202 | 0.1043 |
| 5.1 | 9.1 | 0.0362 | 0.111 | 0.006 | 0.0211 | 0.1113 |
| 6.1 | 7.1 | 0.0093 | 0.0302 | 0.0014 | 0.0053 | 0.0303 |
| 6.1 | 8.1 | 0.0132 | 0.045 | 0.0019 | 0.0076 | 0.045 |
| 6.1 | 9.1 | 0.0148 | 0.052 | 0.0021 | 0.0085 | 0.052 |
| 7.1 | 8.1 | 0.0039 | 0.0148 | 0.0005 | 0.0022 | 0.0148 |
| 7.1 | 9.1 | 0.0056 | 0.0217 | 0.0007 | 0.0031 | 0.0217 |

| 8.1 | 9.1 | 0.0016 | 0.007 | 0.0002 | 0.0009 | 0.007 |

**Table 1** Calculated values of $A(x_1, x_2)$, $B(x_1, x_2)$, $C(x_1, x_2)$, $D(x_1, x_2)$ and $E(x_1, x_2)$.

$$x = \frac{h\nu}{k_B T}.$$

| Quantity | 3-Demensional BB $0.05 \text{ PHz} \leq v \leq 0.19 \text{ PHz}$ | 1-Demensional BB $0.05 \text{ PHz} \leq v \leq 0.19 \text{ PHz}$ |
|---|---|---|
| Total energy density | $2.25 \times 10^{-7}$ J m$^{-3}$ | $9.82 \times 10^{-19}$ J m$^{-1}$ |
| Stefan-Boltzmann law | 16.902 W m$^{-2}$ | $7.36 \times 10^{-11}$ W |
| Number density of photons | $5.91 \times 10^{12}$ m$^{-3}$ | $1.09 \times 10^{3}$ m$^{-1}$ |
| Helmholtz free energy density | $-2.43 \times 10^{-8}$ J m$^{-3}$ | $-1.08 \times 10^{-19}$ J m$^{-1}$ |
| Entropy density | $8.38 \times 10^{-10}$ J m$^{-3}$ K$^{-1}$ | $3.66 \times 10^{-21}$ J m$^{-1}$ K$^{-1}$ |
| Pressure | $2.43 \times 10^{-8}$ J m$^{-3}$ | $1.08 \times 10^{-19}$ J m$^{-1}$ |
| Heat capacity at constant volume | $2.51 \times 10^{-9}$ J m$^{-3}$ K$^{-1}$ | $3.02 \times 10^{-20}$ J m$^{-1}$ K$^{-1}$ |

**Table 2** Calculated values of the thermal radiative and thermodynamic functions of 3 and 1-dimensional black-body radiation in the frequency range 0.05 – 0.19 PHz at $T = 298$ K.

| Quantity | 3-Demensional BB 0.05 PHz $\leq v \leq$ 0.19 PHz | 1-Demensional BB 0.05 PHz $\leq v \leq$ 0.19 PHz |
|---|---|---|
| Total energy density | $1.53 \times 10^{-3}$ J m$^{-3}$ | $2.87 \times 10^{-15}$ J m$^{-1}$ |
| Stefan-Boltzmann law | $1.15 \times 10^{5}$ W m$^{-2}$ | $2.15 \times 10^{-7}$ W |
| Number density of photons | $2.50 \times 10^{16}$ m$^{-3}$ | $2.54 \times 10^{5}$ m$^{-1}$ |
| Helmholtz free energy density | $-4.28 \times 10^{-4}$ J m$^{-3}$ | $-9.78 \times 10^{-16}$ J m$^{-1}$ |
| Entropy density | $1.54 \times 10^{-6}$ J m$^{-3}$ K$^{-1}$ | $3.03 \times 10^{-18}$ J m$^{-1}$ K$^{-1}$ |
| Pressure | $4.28 \times 10^{-4}$ J m$^{-3}$ | $9.78 \times 10^{-16}$ J m$^{-1}$ |
| Heat capacity at constant volume | $4.60 \times 10^{-6}$ J m$^{-3}$ K$^{-1}$ | $7.43 \times 10^{-18}$ J m$^{-1}$ K$^{-1}$ |

**Table 3** Calculated values of the thermal radiative and thermodynamic functions of 3 and 1-dimensional black-body radiation in the frequency range 0.05 – 0.19 PHz at $T$ = 1273 K.

| Quantity | 3-Demensional BB $0.05\ \text{PHz} \leq v \leq 0.19\ \text{PHz}$ | 1-Demensional BB $0.05\ \text{PHz} \leq v \leq 0.19\ \text{PHz}$ |
|---|---|---|
| Total energy density | $1.13 \times 10^{-2}$ J m$^{-3}$ | $1.47 \times 10^{-14}$ J m$^{-1}$ |
| Stefan-Boltzmann law | $8.47 \times 10^{5}$ W m$^{-2}$ | $1.11 \times 10^{-6}$ W |
| Number density of photons | $1.53 \times 10^{17}$ m$^{-3}$ | $4.87 \times 10^{5}$ m$^{-1}$ |
| Helmholtz free energy density | $-4.47 \times 10^{-3}$ J m$^{-3}$ | $-7.26 \times 10^{-15}$ J m$^{-1}$ |
| Entropy density | $6.93 \times 10^{-6}$ J m$^{-3}$ K$^{-1}$ | $9.68 \times 10^{-18}$ J m$^{-1}$ K$^{-1}$ |
| Pressure | $4.47 \times 10^{-3}$ J m$^{-3}$ | $7.26 \times 10^{-15}$ J m$^{-1}$ |
| Heat capacity at constant volume | $2.06 \times 10^{-5}$ J m$^{-3}$ K$^{-1}$ | $1.56 \times 10^{-17}$ J m$^{-1}$ K$^{-1}$ |

**Table 4** Calculated values of the thermal radiative and thermodynamic functions of 3 and 1-dimensional black-body radiation in the frequency range 0.05 – 0.19 PHz at $T = 2273$ K.